\documentclass{iopart}

\usepackage{latexsym,graphicx}
\begin{document}

\title[Energy spectrum of electromagnetic normal modes in dissipative
media]{Energy spectrum of electromagnetic normal modes in dissipative
media: modes between two metal half spaces}

\author{Bo E. Sernelius}

\address{Department of Physics, Chemistry, and Biology, Link{\"o}ping
University, SE-581 83 Link{\"o}ping, Sweden}
\ead{bos@ifm.liu.se}
\begin{abstract}
 The energy spectrum of the electromagnetic normal modes plays a central
 role in the theory of the van der Waals and Casimir interaction. In
 dissipative media it appears as if there are distinct normal modes with
 complex valued energies. The summation of the zero-point energies of
 these modes render a complex valued result. Using the contour
 integration, resulting from the use of the generalized argument principle,
 gives a real valued and different result. We resolve this contradiction
 and show that the spectrum of true normal modes form a continuum with real
 frequencies. We illustrate the problem in connection with the van der
 Waals interaction between two metal half spaces.

\end{abstract}



\section{Introduction}

 The interaction energy in a system may be expressed in terms of the energy
 shifts of the electromagnetic normal modes of the system \cite {SerWiley};
 the longitudinal bulk modes can be used to find, e.g., the polaron energy in
 a polar semiconductor or the exchange-correlation energy in a metal; the
 transverse modes to find the Lamb shift \cite{Feynman}; the surface modes
 to find the van der Waals interaction between objects; the vacuum modes to
 find the Casimir \cite{Casimir} interaction. In all these cases the
 energy, or frequency, of a normal mode is found as a solution to an
 equation of the type

\begin{equation}
\label{ModeCond}
f\left( \omega \right)=0,
\end{equation}
the condition for having a mode. The function $f\left( \omega \right)$,
which involves the dielectric properties of the system, is often obtained
as a determinant of a matrix. The solution to equation (\ref{ModeCond})
approaches a pole of the function $f\left( \omega \right)$ when the
interaction in the system is gradually turned off. When the interaction is
turned on the zeros move away from the poles and $\hbar /2$ times this
shift in frequency is the contribution to the interaction energy of this
particular mode; the interaction energy is the change in total zero-point
energy when the interaction in the system is turned on.

It is often straight forward to use this sum-over-modes approach, or
mode-summation method, to obtain the interaction energy as the following
sum over zeros and poles of the function $f\left( \omega \right)$:
\begin{equation}
\label{ModeSum}
E={\hbar \over 2}\sum\limits_i {\left( {\omega _{0,i}-\omega _{\infty ,i}}
\right)}.
\end{equation}
However, sometimes it is not possible to find an explicit solution to
equation (\ref{ModeCond}) and sometimes the poles and zeros form a
continuum. In those cases the result may be obtained with the so called
generalized argument principle \cite{SerWiley},
\begin{equation}
\label{ArgPrinc}
E={1 \over {2\pi i}}\oint {dz{\hbar \over 2}z{d \over {dz}}\ln f\left( z
\right)},
\end{equation}
where the integration is performed along a contour including the poles and
zeros in the right half of the complex frequency plane. The integration
should be performed in the positive sense, i.e. in the
counter-clockwise direction.

The specific problem of the force between two semi-infinite dielectric
slabs separated by a dielectric layer has been solved with different
approaches; in Lifshitz' \cite{Lifshitz} very complicated theory the
dielectrics are characterized by randomly fluctuating sources as demanded
by the fluctuation-dissipation theorem; Schwinger et al. \cite{Schwinger}
derived the force using Schwinger's source theory, where "the vacuum is
regarded as truly a state with all physical properties equal to zero"; van
Kampen et al. \cite{vanKamp} applied equations (\ref{ModeSum}) and
(\ref{ArgPrinc}) to the surface modes of the geometry to find the
interaction energy and force.

All these approaches, that appear to be quite different, give one and the
same result. Milonni and Shih \cite{Milonni} made a study, on how these
theories are related, based on conventional QED. The result in equation
(\ref{ArgPrinc}) is consistent with what one arrives at from many-particle
theory where the focus is put on the interacting particles in the system,
not the electromagnetic normal modes; there is no explicit reference to
zero-point energies. This is demonstrated in detail in \cite{SerWiley} in
the case of the exchange-correlation energy of a metal. In \cite{SerBjo}
the van der Waals and Casimir forces between two quantum wells were derived
both in terms of the zero-point energy of the normal modes and as the
result from correlation energy; both approaches produced the same result. 
Thus, there are different complementary approaches to the interactions in a
system.

In the present work we are concerned with the sum-over-modes approach in
presence of dissipation in the system. Van Kampen et al. \cite{vanKamp}
considered non-dissipative materials, only, with real-valued dielectric
functions. In the case of dissipative materials the sum-over-modes
approach runs into yet another problem. The straight forward solution of
equation (\ref{ModeCond}) produces complex-valued frequencies or energies,
signalling that the normal modes are no longer stable -- the modes decay. 
If this were the case they would no longer be true normal modes. The
interaction energy in equation (\ref{ModeSum}) becomes complex valued. 
Using instead equation (\ref{ArgPrinc}) produces a real-valued interaction
energy. This situation has made many researchers confused and led to
believe that there is something fundamentally wrong with the mode-summation
method. With the present work we will try to resolve this confusion and
show that the result of equation (\ref{ArgPrinc}) is the correct result in
the case of dissipation; the complex-valued result of equation
(\ref{ModeSum}) is not correct. In an earlier brief report
\cite{SerBrief} we have illustrated the problem by considering the 
longitudinal bulk modes in a metal. In the present work we study the modes
associated with the van der Waals interaction between two metal half
spaces.

In section \ref{Analytical} we discuss the analytical properties of the
dielectric function and show that its zeros and poles are all located at 
the real frequency axis. In section \ref{Example} we apply the different 
formalisms to the specific example viz. the van der Waals interaction between 
two metal half spaces. We introduce a modified mode-summation method that 
works reasonably well in section \ref{ModeSum2} and finish with summary and 
conclusions in section \ref{Summary}.

\section{\label{Analytical}Analytical properties of a dielectric function}

In our illustrating example we will use metals and represent the dielectric 
function with one of the Drude type,
\begin{equation}
\label{DrudeFunc}
\varepsilon \left( \omega \right)=1-{{\omega _{pl}^2} \mathord{\left/
{\vphantom {{\omega _{pl}^2} {\omega \left( {\omega +i\eta } \right)}}}
\right. \kern-\nulldelimiterspace} {\omega \left( {\omega +i\eta }
\right)}}.
\end{equation}
We let the parameter $\eta$ be a positive real-valued constant; $\eta$,
which is the result of electron scattering against impurities or other
defects, is really $\omega$-dependent and complex valued; the approximation
we use is good for a metal in the low momentum limit, for frequencies below
the plasma frequency, $\omega_{pl}$. 

Before we continue let us discuss the general analytical properties of a
dielectric function. The physical dielectric function, the one that can be
measured in experiments, exists on the real frequency axis, only. It is
retarded, which means that it obeys causality. In theoretical treatments
one obtains a function that is analytical in the whole complex frequency
plane except on the real axis, where all the poles are situated. To obtain
the retarded version one either shifts all the poles downwards to an
infinitesimal distance below the real axis and perform the calculation on
the real axis. Alternatively one lets the poles stay put and perform the
calculation just above the axis. For the discussion in this work it is
better to use this last method. There are other versions of the function,
advanced, timeordered and antitimeordered. The timeordered is often used
in many-body calculations since it allows some very useful theorems to be
used. With this version one calculates the function just above the
positive real axis and below the negative real axis. All different
versions are identical everywhere except at the real frequency axis. From
now on, if not stated otherwise, when we discuss the dielectric function we
mean the function with its poles on the real axis. The function has the
properties $\varepsilon \left( { - \omega } \right) = \varepsilon \left(
\omega \right)$ and $ \varepsilon \left( {\omega ^{*}} \right) =
\varepsilon \left( \omega \right)^{*}$. From these equations follows that
the relation between the different forms of the function in the lower and
upper half plane is $\varepsilon _l \left( \omega \right) = \varepsilon _u
\left( {\omega ^{*}}\right)^{*}$. For the full dielectric function with
frequency dependent $\eta$ these two analytical expressions are the same
but not when $\eta$ is treated as a constant. Then we have $
\varepsilon _u \left( \omega \right) = 1 - {{\omega _{pl}^2 } \mathord{\left/
 {\vphantom {{\omega _{pl}^2 } {\omega \left( {\omega + i\eta } \right)}}}
 \right. \kern-\nulldelimiterspace} {\omega \left( {\omega + i\eta }
 \right)}} $ and $
\varepsilon _l \left( \omega \right) = 1 - {{\omega _{pl}^2 } \mathord{\left/
 {\vphantom {{\omega _{pl}^2 } {\omega \left( {\omega - i\eta } \right)}}} \right.
 \kern-\nulldelimiterspace} {\omega \left( {\omega - i\eta } \right)}}
$, respectively. In the full treatment we have
\begin{equation}
i\eta \left( {\omega ^{*}} \right) = \left[ {i\eta \left( \omega \right)}
\right]^{*} = - i\left[ {\eta \left( \omega \right)} \right]^{*} =
 - i{\mathop{\rm Re}\nolimits} \left[ {\eta \left( \omega \right)} \right]
 - {\mathop{\rm Im}\nolimits} \left[ {\eta \left( \omega \right)} \right],
\end{equation}
and we may identify
\begin{equation}
{\mathop{\rm Re}\nolimits} \left[ {\eta \left( {\omega ^{*}} \right)} \right]
= - {\mathop{\rm Re}\nolimits} \left[ {\eta \left( \omega \right)}
\right];\,\,{\mathop{\rm Im}\nolimits} \left[ {\eta \left( {\omega ^{*}}
\right)} \right] = {\mathop{\rm Im}\nolimits} \left[ {\eta \left( \omega
\right)} \right].
\end{equation}
We see that the dominating real part changes sign when we cross the real
axis while the imaginary part does not.

As an illustration, let us look at the expression for the dielectric
function of an impure metal in the so called generalized Drude approach
\cite{SerPRB40}. Let $n_i$, $n$, $S(\bf q)$, and $\omega
_0(\bf q)$ be the density of impurities, electron density, structure factor
for the impurities and impurity potential, respectively. Then the
dielectric function in the small momentum limit is
\begin{equation}
\label{GenDrude}
\fl
 \varepsilon \left( \omega \right) = 1 - \frac{{\omega _{pl}^2 }}{{\omega
 \left\{ {\omega + \frac{{n_i }}{{24\pi e^2 m_e \omega n\Omega
 }}\sum\limits_{\bf{q}} {S\left( {\bf{q}} \right)q^4 \left| {\omega _0
 \left( {\bf{q}} \right)} \right|^2 \left[ {\varepsilon ^{ - 1} \left(
 {{\bf{q}},\omega } \right) - \varepsilon ^{ - 1} \left( {{\bf{q}},0}
 \right)} \right]} } \right\}}},
\end{equation}
where $\Omega$ is volume of the system. The second term within the
parentheses in the denominator is $i\eta$. For a large system the
summation over momentum is usually replaced by an integral over a
continuous momentum variable. Let us now keep the discrete summation. The
derivation of the dielectric function of the pure metal at finite momentum
also contains a discrete summation, now over the electron momentum. The
function in RPA (Random Phase Approximation) is
\begin{equation}
\begin{array}{l}
\fl
 \varepsilon ({\bf{q}},\omega ) = 1 + \left( {{{v_q } \mathord{\left/
 {\vphantom {{v_q } \Omega }} \right. \kern-\nulldelimiterspace} \Omega }}
 \right)\sum\limits_{{\bf{k}},\sigma } {n({\bf{k}})\left[ {1 - n\left(
 {{\bf{k}} + {\bf{q}}} \right)} \right]} \\
\times \left\{ {\left[ {\hbar \omega + \left( {\varepsilon _{{\bf{k}} + {\bf{q}}}
 - \varepsilon _{\bf{k}} } \right)} \right]^{ - 1} - \left[ {\hbar \omega -
 \left( {\varepsilon _{{\bf{k}} + {\bf{q}}} - \varepsilon _{\bf{k}} }
 \right)} \right]^{ - 1} } \right\}, \\ \end{array}
\end{equation}
where $n(\bf k)$ is the Fermi-Dirac occupation number. We see that, if we
make the calculation just off the real frequency axis, the imaginary part
consists of a sum of $\delta$-functions, infinitesimally spaced when the
volume of the system goes to infinity. These form the single particle
continuum in the $\omega q$-plane. The real part passes through zero
between each neighbouring pairs of $\delta$-functions. When the volume goes
to infinity one can replace the summation by an integral. The imaginary
part then turns into a smooth continuous function and the real part does no
longer pass through zero inside the continuum. When one wants to find the
zeros and poles of the function $\varepsilon (\omega)$, in equation 
(\ref{GenDrude}), one should keep the discrete summations everywhere. Then
one realizes that this function, also, has its poles and zeros on the real
axis and that they are in the form of a continuum. 

If we now instead let all summations turn into integrals the parameter
$\eta$ is a smooth complex valued function of frequency. However its real
part dominates and is almost constant for frequencies below the plasma
frequency. From the expression in equation (\ref{GenDrude}) we can easily
verify that the real part of $\eta$ changes sign when one crosses the real
axis. For higher frequencies the contribution to the interaction energy
quickly drops off so equation (\ref{DrudeFunc}) is good enough for our purpose
here.

\section{\label{Example}An illustrating example}

 As an illustration we have chosen the van der Waals interaction between
 two metal half spaces separated by the distance $d$. The modes in this
 system are characterized by the 2D (two-dimensional) wave vector $\bf k$. 
 There are several modes for each wave vector. To find the modes of wave
 vector $\bf k$ we let $f(\omega)$ in equation (\ref{ModeCond}) be \cite
 {SerWiley}
\begin{eqnarray}
\label{ModeCondk}
\fl
\nonumber f_{\bf{k}} \left( \omega \right) = \left[ {\varepsilon \left( \omega
 \right) + 1} \right]^2 - e^{ - 2kd} \left[ {\varepsilon \left( \omega
 \right) - 1} \right]^2 = \left\{ {\left[ {\varepsilon \left( \omega
 \right) + 1} \right] - e^{ - kd} \left[ {\varepsilon \left( \omega \right)
 - 1} \right]} \right\} \\ \times \left\{ {\left[ {\varepsilon \left(
 \omega \right) + 1} \right] + e^{ - kd} \left[ {\varepsilon \left( \omega
 \right) - 1} \right]} \right\} = f_{\bf{k}}^1 \left( \omega
 \right)f_{\bf{k}}^2 \left( \omega \right).
\end{eqnarray}
Equation (\ref {ModeCond}) then gives us four zeros, The two in the right half plane 
are
\begin{eqnarray}
\label{ModeDisp}
\fl
\omega _{\bf{k}} = \left\{ \begin{array}{l}
 \omega _{pl} \sqrt {\left[ {1 + \coth \left( {{{kd} \mathord{\left/
 {\vphantom {{kd} 2}} \right. \kern-\nulldelimiterspace} 2}} \right)}
 \right]^{ - 1} - \left( {{\eta \mathord{\left/ {\vphantom {\eta {2\omega
 _{pl} }}} \right. \kern-\nulldelimiterspace} {2\omega _{pl} }}} \right)^2
 } - {{i\eta } \mathord{\left/ {\vphantom {{i\eta } 2}} \right. 
 \kern-\nulldelimiterspace} 2};\quad f_{\bf{k}}^1 \left( \omega \right) = 0
 \\ \omega _{pl} \sqrt {\left[ {1 + \tanh \left( {{{kd} \mathord{\left/
 {\vphantom {{kd} 2}} \right. \kern-\nulldelimiterspace} 2}} \right)}
 \right]^{ - 1} - \left( {{\eta \mathord{\left/ {\vphantom {\eta {2\omega
 _{pl} }}} \right. \kern-\nulldelimiterspace} {2\omega _{pl} }}} \right)^2
 } - {{i\eta } \mathord{\left/ {\vphantom {{i\eta } 2}} \right. 
 \kern-\nulldelimiterspace} 2};\quad f_{\bf{k}}^2 \left( \omega \right) = 0
 \\ \end{array} \right.
\end{eqnarray}
Thus the zeros are below the real frequency axis. The problem is that the
expression in equation (\ref{DrudeFunc}) for the dielectric function is
only valid above the real axis. Below the real axis $\eta$ has the
opposite sign. The zero is very illusive. If one approaches the zero from
the upper half plane it makes a jump to the upper half plane when one
crosses the real axis. This clearly shows that it is not safe to use
equation (\ref{ModeSum}), directly. If plotted the real part of the mode
energies from equation (\ref{ModeDisp}) form two branches of modes, one
acoustical (lower) and one optical (upper). The lower (upper) branch comes
from putting $f_{\bf{k}}^1 = 0$ ($f_{\bf{k}}^2 = 0$). A first naive guess
is that a good approximation of the interaction energy would be to just sum
the real part of the zero-point energies for the modes. This might seem to
be a good and simple short cut to an approximate result. However, as we 
will see later, it turns out not to be such a good idea.

Let us now instead use equation (\ref{ArgPrinc}) to find the energy
contribution from each 2D wave vector. The total interaction energy per
unit area is
\begin{equation}
E = \frac{1}{\Omega }\sum\limits_{\bf{k}} {E_{\bf{k}} = } \frac{1}{\Omega
}\sum\limits_{\bf{k}} {\frac{1}{{2\pi i}}\oint {dz\frac{\hbar
}{2}z\frac{d}{{dz}}\ln f_{\bf{k}} \left( z \right)} },
\end{equation}
where $\Omega$ is the total area of one half space.  In an actual
calculation we let this area go to infinity and the summation over $\bf k$
turns into an integration.  We first choose our contour to encircle the
positive real axis.  The contour then consists of two parts; an integration
from zero to plus infinity performed just below the real axis; an
integration from plus infinity to zero performed just above the axis.  We
may first perform an integration by parts in both contributions and end up
with

\begin{equation}
E_{\bf{k}} = - \frac{1}{{2\pi i}}\frac{\hbar }{2}\oint {dz\ln f_{\bf{k}}
\left( z \right)}.
\end{equation}
Then the integration below and above the axis are combined into one integral
\begin{equation}
\label{IntRAxis}
\fl
 \nonumber E_{\bf{k}} = - \frac{1}{i}\frac{\hbar }{2}\int\limits_0^\infty
 {\frac{{d\omega }}{{2\pi }}ln} \frac{{f_{\bf{k}} \left( \omega \right)^*
 }}{{f_{\bf{k}} \left( \omega \right)}} = \sum\limits_{l = 1}^2
 {\frac{\hbar }{2}\int\limits_0^\infty {\frac{{d\omega }}{{2\pi }}2tan^{ -
 1} \left\{ {{{Im\left[ {f_{\bf{k}}^l \left( \omega \right)} \right]}
 \mathord{\left/ {\vphantom {{Im\left[ {f_{\bf{k}}^l \left( \omega \right)}
 \right]} {Re\left[ {f_{\bf{k}}^l \left( \omega \right)} \right]}}} \right. 
 \kern-\nulldelimiterspace} {Re\left[ {f_{\bf{k}}^l \left( \omega \right)}
 \right]}}} \right\}} },
\end{equation}
where the function $tan^{-1}$ is taken from the branch where $0 \le \tan ^{
- 1} \le \pi $.

\begin{figure}
\center
\includegraphics{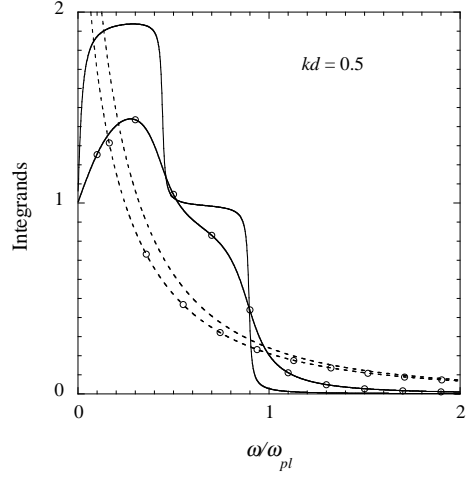}
\caption{The solid and dashed curves are the integrands of equations 
(\ref{IntF}) and
(\ref{IntG}), respectively. The curves with circles are for $x = 0.1$ and
the others are for $x = 0.01$.}
\label{figu1}
\end{figure}

Let us now instead deform the integration contour into a semicircle in the
right half plane with the center of the circle at the origin and the
straight part parallel with and just to the right of the imaginary
frequency axis,
\begin{equation}
\label{IntIAxis}
E_{\bf{k}} = \frac{\hbar }{2}\int\limits_0^\infty {\frac{{d\omega }}{{2\pi
}}2ln} \frac{{f_{\bf{k}} ^{' }\left( \omega \right)}}{4} = \frac{\hbar
}{2}\int\limits_0^\infty {\frac{{d\omega }}{{2\pi }}2ln} \frac{{f_{\bf{k}}
\left( {i\omega } \right)}}{4}.
\end{equation}
The integration along the curved part of the contour vanishes when the
radius goes to infinity. The four in the denominator takes care of this. 
The zeros and poles of the function $f_{\bf{k}} \left( \omega \right)$ in
equation (\ref{ModeCondk}) do not change from the division by a factor of four. 
Thus we end up with two integrals for the same thing; one along the real
frequency axis; one along the imaginary axis. Now, we may simplify the
integrands by letting $ x = {{\left( {{\eta \mathord{\left/ {\vphantom
{\eta 2}} \right.
\kern-\nulldelimiterspace} 2}} \right)} \mathord{\left/ {\vphantom {{\left(
{{\eta \mathord{\left/ {\vphantom {\eta 2}} \right. 
\kern-\nulldelimiterspace} 2}} \right)} {\omega _{pl} }}} \right. 
\kern-\nulldelimiterspace} {\omega _{pl} }} $ and expressing the frequency
in terms of the plasma frequency. Equations (\ref{IntRAxis}) and
(\ref{IntIAxis}) then reduce into
\begin{eqnarray}
\label{IntF}
\fl
\frac{{E_{\bf{k}} }}{{{{\hbar \omega _{pl} } \mathord{\left/
 {\vphantom {{\hbar \omega _{pl} } 2}} \right. \kern-\nulldelimiterspace}
 2}}} = \int\limits_0^\infty {\frac{{d\omega }}{{2\pi }}2\left\{ {\tan ^{
 - 1} \left[ {\frac{{2x\left( {1 - e^{ - kd} } \right)}}{{\omega \left(
 {2\omega ^2 + 8x^2 - 1 + e^{ - kd} } \right)}}} \right]} \right.}
 \nonumber \\ + \left. {\tan ^{ - 1} \left[ {\frac{{2x\left( {1 + e^{ -
 kd} } \right)}}{{\omega \left( {2\omega ^2 + 8x^2 - 1 - e^{ - kd} }
 \right)}}} \right]} \right\} = \int\limits_0^\infty {d\omega F\left(
 \omega \right)}, 
\end{eqnarray}
and
\begin{equation}
\label{IntG}
\fl
\frac{{E_{\bf{k}} }}{{{{\hbar \omega _{pl} } \mathord{\left/
 {\vphantom {{\hbar \omega _{pl} } 2}} \right. \kern-\nulldelimiterspace}
 2}}} = \int\limits_0^\infty {\frac{{d\omega }}{{2\pi }}2ln}
 \frac{{\left[ {2 + {1 \mathord{\left/ {\vphantom {1 {\omega \left( {\omega
 + 2x} \right)}}} \right. \kern-\nulldelimiterspace} {\omega \left(
 {\omega + 2x} \right)}}} \right]^2 - e^{ - 2kd} \left[ {{1 \mathord{\left/
 {\vphantom {1 {\omega \left( {\omega + 2x} \right)}}} \right. 
 \kern-\nulldelimiterspace} {\omega \left( {\omega + 2x} \right)}}}
 \right]^2 }}{4} = \int\limits_0^\infty {d\omega G\left(
 \omega \right)},
 \end{equation}
respectively. These two integrands $F(\omega)$ and $G(\omega)$ are shown
in figure \ref{figu1} for a wave vector with $kd = 0.5$, for the two
parameter choices $x = 0.1$ and $0.01$.

\begin{figure}
\center
\includegraphics{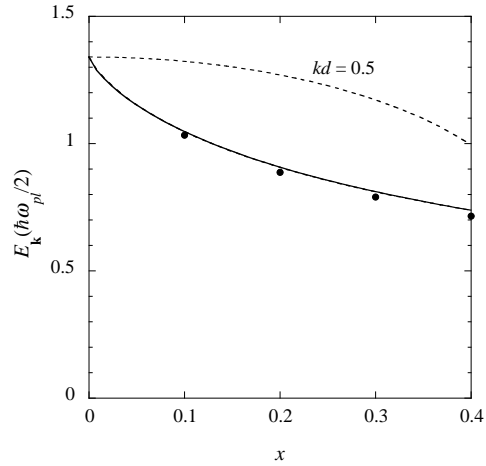}
\caption{The contribution to the interaction energy from one wave vector
with $kd = 0.5.$ The solid curve is the exact result as calculated from
either equation (\ref{IntF}) or (\ref{IntG}). The dashed curve is from
neglecting the imaginary parts of the poles and zeros. The circles are the
result from using equation (\ref{Summation}) and the parameters given in
the text.}
\label{figu2}
\end{figure}

In figure \ref{figu2} we present the results from the different approaches. 
The solid curve is the exact result from equation (\ref{IntF}) or
(\ref{IntG}) and the dashed curve shows the short cut from just dropping 
the imaginary parts of the mode energies. The circles are the
result from yet another approximation that we will discuss in section
\ref{ModeSum2}. The fact that equations (\ref{IntF}) and
(\ref{IntG}) produce the same result proves that the poles and zeros are 
all on the real axis.

\section{\label{ModeSum2}A mode summation method that works}

In the preceding section we found that the naive use of the mode-summation
method failed. The question is whether the mode-summation method can be
used at all in the case of dissipative materials. To investigate this we
make a Lehman representation of the dielectric function,
\begin{equation}
\label{Lehman}
\fl
 \varepsilon \left( \omega \right) = 1 + \int\limits_{ - \infty }^\infty
 {\frac{{d\omega ^{'}}}{{2\pi }}} \frac{{ - 2{\mathop{\rm Im}\nolimits}
 \varepsilon \left( {\omega ^{'} } \right)}}{{\omega - \omega ^{'} }}
 \nonumber = 1 - \int\limits_0^\infty {\frac{{d\omega ^{'} }}{{2\pi }}}
 \frac{8x}{{\left[ {\left( {{\omega \mathord{\left/ {\vphantom {\omega
 {\omega _{pl} }}} \right. \kern-\nulldelimiterspace} {\omega _{pl} }}}
 \right)^2 - \left( {\omega ^{'}} \right)^2 } \right]\left[ {\left( {\omega
 ^{'} } \right)^2 + 4x^2 } \right]}}.
\end{equation}
The integral over frequency can be viewed as a limit of discrete frequency
summations where the step size goes towards zero. In each summation the
dielectric function has its poles and zeros on the axis. These points come
closer and closer when we take the limit. The integrations over frequency
and momentum should always be viewed as limits of discrete summations since
the system is always finite in size and possible energy and momentum
transfers are discrete.

Now if we approximate the integral of equation (\ref{Lehman}) with a discrete 
summation, 
\begin{eqnarray}
\fl
 \varepsilon \left( \omega \right) \approx 1 - \frac{{8x}}{{2\pi
 }}\frac{{\omega ^{'} _{\max } }}{{i_{\max } }}\sum\limits_{i = 1}^{i_{\max
 } } {\left[ {\left( {{\omega \mathord{\left/ {\vphantom {\omega {\omega
 _{pl} }}} \right. \kern-\nulldelimiterspace} {\omega _{pl} }}} \right)^2
 - \left( {\omega _i ^{'} } \right)^2 } \right]^{ - 1} \left[ {\left(
 {\omega _i ^{'} } \right)^2 + 4x^2 } \right]^{ - 1} },
\label{Summation}
\end{eqnarray}
 where $ \omega _i ^{'} = \left( {i - {1 \mathord{\left/ {\vphantom {1 2}}
 \right. \kern-\nulldelimiterspace} 2}} \right){{\omega ^{'} _{\max } }
 \mathord{\left/ {\vphantom {{\omega ^{'} _{\max } } {i_{\max } }}} \right. 
 \kern-\nulldelimiterspace} {i_{\max } }}$, the dielectric function has a
 finite number of poles and zeros. Using this function in the expressions
 for the two factors $f_{\bf{k}}^1 \left( \omega
\right)$ and $f_{\bf{k}}^2 \left( \omega\right)$ of equation (\ref{ModeCondk})
means that also these factors each have the same number of poles and zeros. 
Using equation (\ref{ModeSum}) where the sum runs over these poles and
zeros gives us an approximate result for $E_{\bf{k}}$, the contribution to
the interaction energy from mode $\bf{k}$. The result becomes
asymptotically the exact result when we let $\omega ^{'} _{\max } $ and
$i_{\max } $ go towards infinity. In figure \ref{figu2} the first (last)
two circles were calculated including 50 equidistant poles in the region
below $ {\rm{2}}\omega _{{\rm{pl}}}$ ($ {\rm{3}}\omega _{{\rm{pl}}}$), i.e.
$\omega ^{'} _{\max } = 2(3)$ and $i_{\max } = 50(50)$. Thus we have to
find the frequency of 50 poles and 50 zeros. This is feasible and the
result is in much better agreement with the exact result than what one
obtains when just neglecting the imaginary part of the zeros and poles.

\begin{figure}
\center
\includegraphics{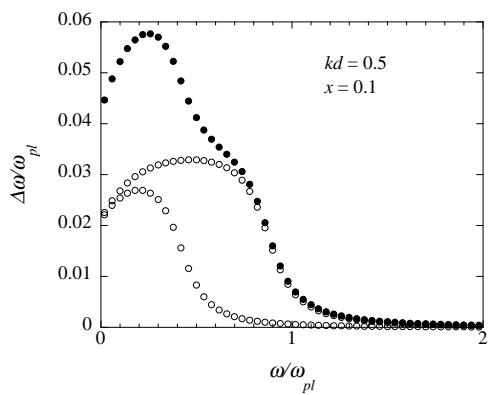}
\caption{The energy shifts of the zeros away from the poles when the
interaction is turned on.
The open circles are the shifts of the modes corresponding to one of the 
two branch types. The filled circles is the combined effect. See the text 
for more details}
\label{figu3}
\end{figure}

In this treatment all poles and zeros are real-valued. We now have one
acoustical and one optical branch in between each pair of neighbouring
poles. When the interaction is turned off the zeros end up at the position
of the lowest of the two poles. When the interaction is turned on they
move up a distance and it is these shifts in frequency or energy that are
their contributions to the interaction energy. In figure \ref{figu3} we
show as open circles these shifts for the acoustical (the curve with
left-most maximum) and optical branches. The combined shifts are shown as
filled circles. We clearly see the resemblance of the filled circles to
the integrand of equation (\ref{IntF}), represented by the solid curve with
circles in figure \ref{figu1}.

\section{\label{Summary}Summary and conclusions}

In summary, we have demonstrated that the energy of the electromagnetic
normal modes in dissipative media are real valued and lead to real-valued
interaction energies and forces. The modes appear to be distinct with
complex valued energies, but they are not. The modes form a continuum on
the real frequency axis and the direct use of the mode summation method is
no longer feasible. The functions appearing in the condition for modes
contain integrals over momentum or frequencies. Since the system is finite
in size these integrals should be considered as discrete summations; the
possible momentum and energy transfers in a finite system is discrete. In
doing so all zeros and poles end up on the real axis. When one takes the
limit when the volume goes to infinity the zeros and poles form a continuum
of points on the real axis. We have demonstrated two different contour
integrations that can be used to find the result; the one we favour most
ends up with an integration along the imaginary frequency axis; the other
involves an integration along the real frequency axis. We have further
demonstrated that one may use an approximation in which this continuum of
modes is replaced by a finite number of distinct modes and obtain
reasonable good results using the mode-summation method. This gives a
great improvement compared to the result from just neglecting the imaginary
parts of the complex valued energies of the ÒapparentÓ modes.

\ack
This research was sponsored by EU within the EC-contract
No:012142-NANOCASE.

\end{document}